OPEN ACCESS Freely available online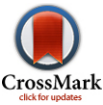

# A Method for Modeling Growth of Organs and Transplants Based on the General Growth Law: Application to the Liver in Dogs and Humans

Yuri K. Shestopaloff[1]*, Ivo F. Sbalzarini[2]

1 Research and Development Lab, Segmentsoft Inc., Toronto, Ontario, Canada, 2 MOSAIC Group, Center of Systems Biology Dresden (CSBD), Max Planck Institute of Molecular Cell Biology and Genetics, Dresden, Germany## Abstract

Understanding biological phenomena requires a systemic approach that incorporates different mechanisms acting on different spatial and temporal scales, since in organisms the workings of all components, such as organelles, cells, and organs interrelate. This inherent interdependency between diverse biological mechanisms, both on the same and on different scales, provides the functioning of an organism capable of maintaining homeostasis and physiological stability through numerous feedback loops. Thus, developing models of organisms and their constituents should be done within the overall systemic context of the studied phenomena. We introduce such a method for modeling growth and regeneration of livers at the organ scale, considering it a part of the overall multi-scale biochemical and biophysical processes of an organism. Our method is based on the earlier discovered general growth law, postulating that any biological growth process comprises a uniquely defined distribution of nutritional resources between maintenance needs and biomass production. Based on this law, we introduce a liver growth model that allows to accurately predicting the growth of liver transplants in dogs and liver grafts in humans. Using this model, we find quantitative growth characteristics, such as the time point when the transition period after surgery is over and the liver resumes normal growth, rates at which hepatocytes are involved in proliferation, etc. We then use the model to determine and quantify otherwise unobservable metabolic properties of livers.**Citation:** Shestopaloff YK, Sbalzarini IF (2014) A Method for Modeling Growth of Organs and Transplants Based on the General Growth Law: Application to the Liver in Dogs and Humans. PLoS ONE 9(6): e99275. doi:10.1371/journal.pone.0099275

**Editor:** Manlio Vinciguerra, University College London, United Kingdom

**Received** March 10, 2014; **Accepted** May 12, 2014; **Published** June 9, 2014

**Copyright:** © 2014 Shestopaloff, Sbalzarini. This is an open-access article distributed under the terms of the Creative Commons Attribution License, which permits unrestricted use, distribution, and reproduction in any medium, provided the original author and source are credited.

**Data Availability:** The authors confirm that all data underlying the findings are fully available without restriction. All used data were published and references to publications are in the article.

**Funding:** The funders had no role in study design, data collection and analysis, decision to publish, or preparation of the manuscript.

**Competing Interests:** There are no competing interests. Although one of authors, Yuri Shestopaloff, works in a commercial company Segmentsoft Inc. as a Director of Research & Development Lab, as you correctly noted, the study has nothing to do with the business activity of the company. This study is an entirely personal undertaking conducted in a spare time. The situation allows me taking unpaid breaks in order to do these studies and visiting other organizations. In particular, part of this study was done at Max Planck Institute of Molecular Cell Biology and Genetics, where I stayed as an invited Visiting Professor (without being paid as well). The topic of the study was not included into any official Institute's research plans, and no funds were allocated for it. So that formally it can be considered as a self-initiated research with no funding. This does not alter our adherence to PLOS ONE policies on sharing data and materials.

* E-mail: shes169@yahoo.ca## Introduction

First we introduce the earlier discovered general growth law and its mathematical representation, the growth equation, and apply it towards modeling growth of livers and liver transplants in dogs and humans (the first article) and finding liver metabolism (the second article). Then, we present a review of presently available models from the perspective of developing a general framework for modeling biological phenomena, and how the general growth law can benefit it. Such a framework, if created correctly, would unite and mutually reinforce available methods and provide directions and guidance for the development of multi-scale models of living organisms and their constituents, such as organs and cells, as well as allow model verification and subsequent refinement. Such a framework is especially important given the many practical problems whose solution requires a transition to *systemic* understanding of living organisms, so that on this well founded basis the following practical applications and methods could be introduced in diverse areas, such as medicine, pharmacology, biology, biotechnology, etc.

Developing such a framework, indeed, became a necessity given the launch of projects aiming at the creation of models of organisms and organs to be used in medicine, pharmacology, biology, evolutionary and developmental studies, etc., such as, e.g., the Virtual Liver Network (VLN) [1], the Recon-2 project on human metabolism [2], the virtual liver project [3], the whole-body model [4], the Physiome Project on cardiac electrophysiology [5], the BlueBrain project on modeling the brain cortex, and others. Such models have different levels of generality addressing certain phenomenological, structural, and organizational aspects. However, since the different mechanisms and systems in organisms closely interrelate, the adequacy and usefulness of models would be improved by including additional mechanisms and components, through interlacing different factors, and unification of methodological approaches based on a general framework.

PLOS ONE | www.plosone.org     1     June 2014 | Volume 9 | Issue 6 | e99275

## Methods

### 1. The general growth law

Growth regulation and modeling growth of cells, organs, and whole organisms is an area of intensive study. Approaches range from studies of biomolecular growth mechanisms and growth factors, to developmental and systems biology methods. For instance, in [6], authors argue that changes during growth, such as progressive decline in proliferation, "results from a genetic program that occurs in multiple organs and involves the down-regulation of a large set of growth-promoting genes." The authors further note that "This program does not appear to be driven simply by time, but rather depends on growth itself, suggesting that the limit on adult body size is imposed by a negative feedback loop." They consider different cellular events that could be involved in cooperatively providing commensurate growth of organs and whole organisms. An important inference is the recognition of the existence of feedback mechanisms between the current *integral* state of a growing organ or an organism (which the authors call "growth itself") and triggering particular growth mechanisms into *cooperative* action.

Reference [7] considers growth hypotheses based on morphogen gradients. They conclude that the growth phenomenon is driven by a *combination* of different factors. A similar view is expressed in [8], which considers growth from a systems-biology perspective. The author suggests that "developing systems devote a considerable amount of cellular machinery to the explicit purpose of control", although he does not specify what this "controlling machine" consists of, or what are the coordinating and managing mechanisms.

All cited articles converge to the conclusion that growth is driven by the cooperative working of many different factors, whose action, besides other possible mechanisms, is regulated by feedback loops. In [6], a guiding mechanism is placed into a "genetic program that occurs in multiple organs", which "depends on growth itself". In other words, the authors assume that the general governance and coordination of biomolecular growth mechanisms resides at the molecular level. Articles [7,8] support similar ideas, that biomolecular mechanisms govern and coordinate the multitude of interacting mechanisms constantly synthesizing and degrading molecules within cells, managing *cooperative* growth of multiple cells, and growth of different organs and systems in the whole organism. These governing molecular mechanisms are assumed in [6] to be a "genetic program" that has to have a complexity on the order of that of the biochemical machinery itself. But still we are unable to explain the coordinated growth of organs and systems within an organism. Such inter-organ genetic regulation would amount to unmanageable complexity and consequently to extreme vulnerability and instability, which we do not observe in nature.

We hence take the view that biochemical mechanisms *execute* operations in such a manner that one operation faithfully follows another, so that there is no need for a run-time scheduler. However, such a sequence of operations had to be evolutionarily developed and organized over a long time. Some researchers assume that such sequences of events are *somehow* stored in DNA. However, the existence of genes does not explain neither how the aforementioned sequence of events has evolved, nor does it give satisfactory answers as to how it unfolds in a particular growth and replication scenario on the cell, organ, and organism levels. So, there should be other than purely genetic mechanisms responsible for growth control. Examples of such views can be found in a seminal work by D'Arcy Thompson [9], and the book [10].

Recent studies [11–17] (the most important and comprehensive work is [17]), discovered that, indeed, such a regulatory mechanism exists at higher-than-molecular scales, which is called the general growth law. This law *universally* operates at scales ranging from cells and cellular components to organs and whole organisms. It is responsible for the evolutionary development of sequentially executed biochemical mechanisms in developing organisms, as well as for unfolding these sequences of events in particular growth and replication scenarios of cells, organs, and organisms. During growth, the general growth law imposes certain constraints on the amount of produced biomass, which accordingly causes changes in composition of biochemical reactions in such a way that the growing entity proceeds through the growth cycle. The same mechanism is also the major player securing *balanced growth* of different organs and systems in an organism [17]. Mathematically, the general growth law is represented by growth equations, which come in different forms depending on the replication and growth scenario.

Previously, the general growth law and the growth equation have been successfully used for studying and modeling the growth and replication mechanisms in unicellular organisms, such as fission yeast and its mutants, amoeba, and *S. cerevisiae* [13,17]. Here, we propose and demonstrate a method for modeling growth of multi-cellular organs. We present mathematical forms of the growth equations for modeling the growth of transplanted livers, liver grafts, and liver remnants in dogs and humans. The purpose of this study is twofold: First we develop a general method, which can be thought of as a methodological framework, that allows to describe, predict, and understand different aspects of the growth of organs, such as finding the rate of growth and its dynamics, the progression in changes of size and geometry, the size (meaning mass and volume) of an organ, identifying certain qualitative phases of growth, etc. Although in this work the proposed method is exemplified by studying the growth of transplanted livers in dogs and humans, the approach itself is of a general nature and can be used in similar applications, including growth of artificial organs, such as kidneys or hearts [18]. The second purpose of this work is to continue the study of the general growth law, including verification aspects. It is also the first time that the general growth law is applied on the organ scale.

Growth and replication of living species are governed by biophysical mechanisms on molecular and higher levels. The general growth law and its mathematical representation, the growth equation, formulate how nutrients are distributed at higher-than-molecular levels and uniquely relate it to metabolic and geometric properties of the growing organism and its constituents, such as organelles, cells, and organs. The general growth law is based on conservation of mass with regard to nutrients, since nutrients are digested in biochemical reactions, for which the law of conservation is valid.

Any living organism is an open system that consumes nutritional resources, which are balanced between two main activities vitally important for any organism: supporting existing biomass, the so-called maintenance resources, and the resources that are used for synthesis of new biomass. This distribution of resources is not arbitrary, but represents a tradeoff that is uniquely defined in *every phase* of growth and replication and on *each spatial* scale. The parameter that mathematically defines this resource division is called the *growth ratio* [17]. It naturally depends on the geometry (shape) of the growing object and, indirectly, on the properties of its biochemical machinery. An optimum distribution of nutritional resources has likely emerged from evolutionary pressures.

As organs grow, more and more resources are required for maintenance, leaving less resources for biomass production, since the nutrient-supplying ability of the environment and the metabolic abilities of the cells are limited. Nutrients, regardless of how they are supplied, are received through the *surface* (of the organ or its blood vessels), while they have to support the functioning of mass in the *volume*. Since volume increases faster than surface area when organisms grow, the nutrient supply per unit volume is *fundamentally* limited. At some point, the amount of nutrients per unit volume decreases to a level that is just sufficient to support maintenance needs, and no nutrients are left for biomass production. This effectively imposes limits on the maximum size of growing organisms and their constituents (besides the specific properties of the biochemical machinery, which, in this regard, plays the role of an execution mechanism). Note that the nutrient concentration in the surrounding environment (for instance, in the blood flow) cannot increase endlessly too, as well as the capacity of an organism or its constituent, such as a cell or a liver, to process nutrients. So, in one way or another, at some point of growth, the amount of nutrients per unit volume will be capped.

The *growth ratio*, which defines the fraction of nutrients that goes to biomass production, depends on the geometric shape of the organ. It is defined as follows: Let us assume that nutrient availability and the biochemical specifics of an organ that receives nutrients through its surface, allow the organ to grow to a maximum volume of $V_{MAX}$ with a maximum surface of $S_{MAX} = S(V_{MAX})$. We define the *dimensionless* relative surface $R_S$ and the relative volume $R_V$ as:

$$R_S = \frac{S(V)}{S(V_{MAX})} \quad (1)$$

$$R_V = \frac{V}{V_{MAX}} \quad (2)$$

Then, the growth ratio $G_R$, which is also dimensionless, is defined as:

$$G_R = \frac{R_S}{R_V} - 1 \quad (3)$$

Although this parameter is described in terms of geometric characteristics, it is closely related to the biochemistry of the organ, since it defines how much nutritional resources are used for growth, in other words, for biomass production, while the rest is used to support the organ's maintenance needs. The particular form of the growth equation depends on the growth scenario. For instance, when nutrients are supplied through the surface, the growth equation can be written as:

$$p_c(X) dV(X,t) = \left( \int_{S(X)} k(X,t) \times dS(X) \right) \times \left( \frac{R_S}{R_V} - 1 \right) dt \quad (4)$$

.

Here, $X$ is the spatial coordinate, $p_c$ is the density of the tissue measured in $kg/m^3$, $t$ is time, $k$ is the specific influx, which is the nutrient influx per unit surface per unit time measured in $kg/(m^2 \times \text{sec})$, $p_c(X) dV(X,t)$ is the change in mass, and $dS(X)$ is the elementary surface area. In case when the specific influx does not depend on the location of an elementary surface area, equation 4 simplifies to

$$p_c(X) dV(X,t) = k(t) \times S \times \left( \frac{R_S}{R_V} - 1 \right) dt \quad (5)$$

where $S$ is the total surface through which nutrients are supplied.

Equation 4 has a simple interpretation: The left-hand side represents the mass increment. The right-hand side represents the total influx through the surface, that is the term $\int_{S(X)} k(X,t) \times dS(X)$, multiplied by the growth ratio $(R_S/R_V - 1)$, so that this product defines the amount of nutrients that is available for biomass production.

Note that the maximum size of a growing organism or an organ can vary, since the size and shape can change during growth depending on many factors, such as nutrient availability, temperature, etc. This *fundamental* property of every growth phenomenon is exactly what the growth equation incorporates through the introduction of a maximum size that can depend on other parameters. This property can be illustrated as follows: It was experimentally found in [19] that cells placed from a nutritionally poor into a nutritionally rich environment grow noticeably bigger. Similarly, suppose an organ started to grow in a nutritionally poor environment so that it is destined to have a smaller final size [17]. If, during growth, the nutritional environment becomes richer, the organ's final size can be larger.

So, unless conditions for the whole growth period are known at the onset of growth, the final size is generally unknown. However, in many instances, the final size of a growing organ is known from prior information, for example when the organ's mass is a well-defined fraction of the mass of the whole organism.

Another approach to finding the maximum size is the following: In an extensive review [20] on tissue growth, the authors note: "A surprising result of this type of modeling (allometric) is that the mass of an organism during its growth process can be predicted based on metabolic processes in its cells.", referring to results obtained in [21]. If we take a look at the growth equation, equation 4, then the "surprising" result finds a rational explanation. According to the growth equation, the rate of biomass synthesis, and consequently the final size, depends on the nutrient influx consumed by the growing organ, which is defined by the *metabolic* abilities of the cells to process nutrients for biomass synthesis and maintenance, which explains the aforementioned result in [21]. In fact, the dependence of an organ's final size on the metabolic properties of its cells and on nutrient availability was *first inferred* from the growth equation, and *then* the search in the literature confirmed this fact.

Mathematically, this property can be expressed as a power law [21], that is "If $y$ is the length scale of the organ, and $x$ is the length scale of the body, they can often be related by a power law of the form $y = x^b$, for constant $a$ and $b$". Reference [22] further advances this result allowing finding the maximum size of a grown organism based on metabolic properties of its cells. The authors proved that "the mass of a wide variety of animal species grew according to the equation $\frac{dm}{dt} = am^{3/4} - bm$, where $a$, $b$ are constants (different for each species), which are dependent on the metabolic characteristics of the cells. The key assumption here is that the metabolic rate $B$ depends on the total body mass $m$ through the power law relation $B \propto m^{3/4}$ which is true for a wide range of biological organisms [23]."

Usage of the fact that the maximum size of a grown multicellular organism depends on the metabolic activity of its cells is facilitated by the growth equation as follows: According to equation 5, the increase of biomass at any given moment is proportional to the nutrient influx $k$, while the functional dependence of the change of nutrient influx for the same organism is similar in a wide range of growth scenarios [13,17]. So, once we know the minimum $k_{min}$ and maximum $k_{max}$ nutrient influxes, corresponding accordingly to the minimum and maximum metabolic rates of the cells and the minimum $m_{min}$ and maximum $m_{max}$ masses of the organism, we can find the maximum mass resulting from influx $k$ as $m = m_{min} + (m_{max} - m_{min})(k - k_{min})/(k_{max} - k_{min})$. Here, we assume that nutrient influxes $k$, $k_{min}$, $k_{max}$ relate to the same phase of growth, let us say to the beginning, and $k_{min} \leq k \leq k_{max}$.

When one does not know how the nutrient influx varies during growth, and consequently how the maximum size changes, the discussed approaches produce approximate values of maximum size. Finding the maximum size is by no means restricted to the described methods. Other considerations and approaches can be used too.

So, the variable maximum size of an organism in the growth equation is just a reflection of the fact that, generally, the maximum final size is a value that is *fundamentally* unknown at the beginning of growth, since the change of growth conditions changes the maximum final size (unless we know how all parameters, which influence the growth, dynamically change during the whole growth period). However, in many instances, when conditions of growth are stable, the maximum size can be predicted with reasonable accuracy for practical purposes.

## 2. Modeling growth of whole livers transplanted from small dogs into large dogs

In [24], the authors measured the growth of whole livers that were transplanted from small dogs, whose weight was $(13.2 \pm 0.4)$ *kg*, to large dogs with weights $(23.7 \pm 0.8)$ *kg*. The control group consisted of dogs with similar weights. In that group, livers from donor dogs with weights $(19.5 \pm 4.5)$ *kg* were transplanted to recipient dogs with weights $(18.7 \pm 4.6)$ *kg*. The goal of the experiments was to find out which factors define the final size of transplanted livers. It turned out that the liver volumes (and accordingly their masses, since the density of a liver is relatively constant) grows to a final size defined by a certain, stable fraction of the overall body mass. In other words, there is no "memory" in a small liver of a small dog that it is small or that it belonged to a small dog. This confirms that the molecular pathways are diligent *reactive executers* of instructions at the cellular level, but not more than that, while the organ's size and geometrical characteristics are defined by other mechanisms. From this study, we consider in more detail two data sets of liver volume over time, from two dogs, for a total observation period of 30 days.

**2.1. Geometrical model of a dog liver.** Dog livers have a shape that is largely defined by the anatomical location and by adjoining organs. We model a dog liver as a partial torus, cut through its plane of symmetry. The parameters defining the torus are: initial distance $d_b$ (index 'b' stands for "beginning", i.e., at the onset of growth) between the torus center and the center of the circle that creates the torus, the initial ($r_b$) and final ($r_e$) radii of the torus at the beginning and at the end of growth, and the number $P$ that defines which fraction of the torus is left (like 2/3 of the total circle). The ends of the torus are capped by two hemispheres, and then the whole shape is cut through its plane of symmetry (see front and side views in Fig. 1).

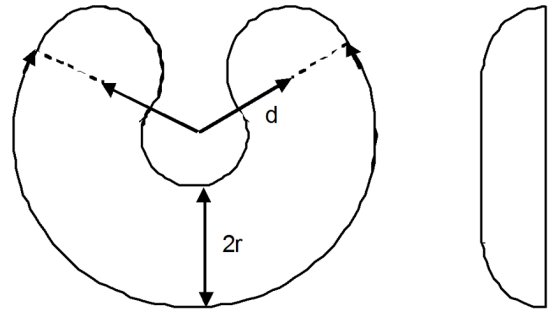

**Figure 1. Front and side views of a partial, sliced torus.** It is used as a geometric model for dog livers.
doi:10.1371/journal.pone.0099275.g001

This shape imitates the growth of a liver that increases proportionally in all dimensions, so that once we know how much the radius of the torus increases (which is defined by the final liver volume), we can find how the size changes of all other dimensions. For example, the distance $d$ is defined through a scaling coefficient $C = r_e/r_b$ as $d_e = Cd_b$. The four parameters ($r_b, r_e, d_b, P$) uniquely define the shape of the dog liver before and after the growth period. We assume that a liver grows proportionally in all dimensions. This is a reasonable, albeit not confirmed assumption, since no indications were made in [24] with regard to the shape of the liver at intermediate phases of growth. Using the notation introduced in Fig. 1, the volume $V$ and the surface $S$ of the liver model are:

$$V(r,d) = P\pi r^2(\pi d + 2/3 r)$$

$$S(r,d) = P\pi r(2\pi d + 4d + r) \quad (6)$$

Accordingly, the relative surface $R_S$ and the relative volume $R_V$, which we need for the growth equation, are:

$$R_S = S(r,d)/S(r_e,d_e) \quad (7)$$

$$R_V = V(r,d)/V(r_e,d_e) \quad (8)$$

For known relative surface and relative volume, the growth ratio $G$ can be found using equation 3.

In order to formulate the growth equation, we have to define the nutrient influx. In a liver, nutrients are supplied through the blood, which flows through the liver as driven by blood pressure. In the portal veins, the blood pressure drops from 130 to 60 mm water. After passing the sinusoids the pressure further drops to 20 mm water. We assume that the amount of nutrients supplied to every position in the liver is the same, and that each unit of volume consumes the same amount of nutrients. Under these assumptions, the growth equation becomes:

$$p\,dV(r,d) = K \times (V(r,d)/V_b) \times \left(\frac{R_S}{R_V} - 1\right) dt \quad (9)$$

where $p$ is the liver density, which we assume to be constant, $K$ is the total influx of nutrients supplied to the liver by the blood per unit volume per unit time, $t$ is time, $V_b$ is the initial liver volume, and $(R_S/R_V - 1)$ is the growth ratio. We normalize $K=1$, since its value defines the unknown time-scaling coefficient.

The assumption of constant density of the liver is well justified, given its anatomical and cytological uniformity [25,26]. Although the composition of the nutrients received by the hepatocytes depends on the location along the sinusoid, the amount of nutrients available per unit volume is assumed to be constant [25]. This is reflected in equation 9 by the multiplier $(V/V_b)$.

We numerically solve equation 9 using the rectangular rule for numerical integration, i.e., by dividing the ranges of the radius $r$ and the distance $d$ into equal intervals and computing the appropriate function values at the centers of the intervals (recall that $r/d = const$). All variables except $t$ in equation 9 depend on volume, so that we collect them on the left-hand side, and integrate over the range of $r$ (and correspondingly $d$) in order to obtain time.

**2.2. Modeling the growth of entire livers in dogs.** The geometric parameters of the growing livers from two dogs, as taken from [24], are summarized in Table 1. This is a *complete* set of parameters required to compute the growth curve using equation 9. Then, we scale the obtained growth curve along the time axis only, in order to adjust the time scale to experimental data. Note that this is *not a data fitting* procedure, because we *first* computed the growth curve, and only *after that* compared it to experimental data. The scaling along the time axis does not change the shape of the growth curve, but rather amounts to identifying time scale $K$ of the observed dynamics.

Due to transition processes occurring in a transplanted liver after resection and surgery, the transplanted liver initially does not grow the same way it would normally grow, and less hepatocytes are involved in replication compared to a normally growing liver. When the liver grows normally, its size increase is described by the growth equation, which represents the evolutionarily optimized growth scenario, securing the shortest growth time. According to [25], which considers hepatectomy with significant resections, gradually all hepatocytes become involved in replication. The authors say: "After tissue loss, residual hepatocytes are activated to proliferate within few hours; hepatocytes proliferation begins at the portal ends of plates …, and successive waves of hepatocytes proliferation ultimately involve virtually all residual hepatocytes. Hepatocytes proliferation is followed sequentially by proliferation of sinusoidal endothelial cells and macrophages, and the other cells of parenchymal matrix". However, towards the end of growth, more and more hepatocytes switch to a quiescent state, since at the end the liver growth decelerates. Another possibly contributing factor could be a slowing of the hepatocyte cell cycle toward the end of growth, but according to [27] switching to a quiescent state is the main cause of growth deceleration. So, although according to [25] there is a relatively long phase of growth when *all* hepatocytes become involved in liver regeneration (in donors and recipients), when the liver is only reduced little by resection, and also towards the end of growth, a noticeable fraction of hepatocytes are in a quiescent state.

In order to identify the time point after which the entire liver grows normally (according to the general growth law), we first assume that the entire liver grows normally from the beginning and compare the so-obtained growth curve to the experimental data. The point after which the curve agrees with the data is the time when the entire liver grows normally. When the resected liver part is significant, then, according to [24], this is also the point after which all hepatocytes are involved in replication. Then, once we know when the entire liver begins to regenerate normally, we can model the preceding phase of partial growth with gradual involvement of hepatocytes.

Using equation 9, we compute growth curves for dog livers and compare the results with experimental data from [24], as shown in Fig. 2. For dog 1 (Fig. 2A) the first experimental point is not taken into account, since it corresponds to the not-yet transplanted liver, when it was weighted right after hepatectomy, while the rest of the experimental points correspond to results obtained by CT scanning. The second point marks the beginning of growth where less hepatocytes than in normal growth are involved in regeneration. As one can see from the graph, this point is off the growth curve computed under the assumption of normal growth.

For the rest of data, the correspondence between the experimental results and the computed growth curve is very good, which is an indication that the proposed approach produces a real dependence. So, the growth equation can serve as an adequate tool for modeling the growth of dog livers.

The computed growth curve for the second dog (Fig. 2B) is also in good agreement with the experimental data after some initial divergence, which indicates the time it takes for the liver to engage in normal growth.

Our model has hence allowed us to identify the time point after which the maximum number of hepatocytes are involved in the regeneration process. This is the point where the growth curve computed under the assumption of normal proliferation starts to agree with the experimental data. For large resections, according to [25], at this stage "virtually all residual hepatocytes" are involved in proliferation. In case of usual hepatectomy, when roughly 30% is removed from the donor liver, according to [25], this will be the point where normal growth resumes (points $V_1$ and $V_2$ in Fig. 2, which we refer to as "joining" points).

**Table 1.** Dimensions of geometric models used for computing dog liver growth, taken from (24).

| Parameter | Dog 1 | Dog 2 |
| --- | --- | --- |
| Initial torus radius $r_b$ | 1 | 1 |
| Final torus radius $r_e$ | 1.291 | 1.45 |
| Initial distance from the center to torus axis $d_b$ | 1.25 | 1.25 |
| Final distance from torus center to torus axis $d_e$ | 1.613 | 1.8125 |
| Fraction of the torus used for modeling | 2/3 | 2/3 |
| Minimum initial volume (cubic centimeters) | 374.28 | 344.778 |
| Final volume (cubic centimeters) | 805.05 | 1049.963 |
| Relative volume (relative to minimum) | 2.1509 | 3.0453 |

doi:10.1371/journal.pone.0099275.t001

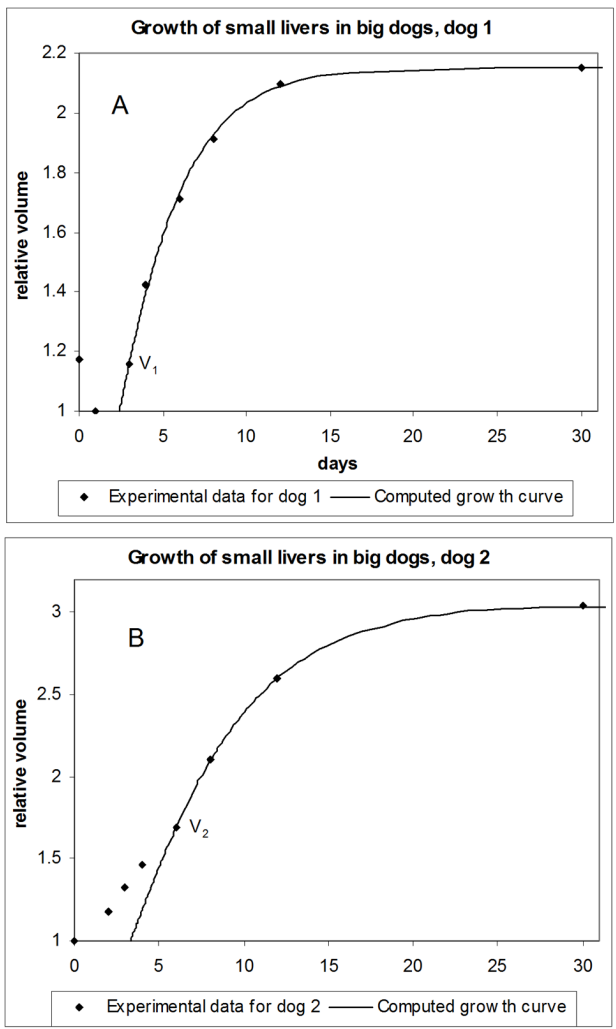

**Figure 2. Growth of small livers in big dogs.** Computed growth curves versus experimental data. A - dog 1; B - dog 2.
doi:10.1371/journal.pone.0099275.g002

### 2.3. Modeling partial growth of livers in dogs.

The phase before the joining points is characterized by partial growth where less hepatocytes than in case of normal growth are contributing to organ increase. As we discussed before, liver transplants do not start growing entirely at once after transplantation, since the liver structure is built sequentially starting from proliferation of hepatocytes. Also, not all hepatocytes are activated for proliferation simultaneously, but are gradually engaged in the proliferation process from the portal ends of plates. However, it is presently unknown what fraction of hepatocytes, relative to normal growth or total volume, are involved in proliferation at the beginning, and when all hepatocytes become involved in regeneration, although these are important characteristics which are directly related to transplantation outcome and recovery process. Our model allows answering these questions.

In order to extend our model to partial growth, given the observed slower growth at the beginning, we assume that a smaller fraction of hepatocytes is initially involved in proliferation. The indication that "*hepatocytes proliferation begins at the portal ends of plates*" [25] means that a fairly large part of the total liver volume is involved in proliferation from the very beginning. For simplicity, we assume that the proliferating hepatocytes are uniformly distributed across the entire organ. If shells of proliferation would exist, as they do for example in other organs in which actively growing areas are located at the periphery, then we would consider such a growing shell and compute the value of the growth ratio for this shell only, and accordingly would apply volumetric characteristics to the shell as well.

During partial growth, we distinguish the growing part of the liver (we call it the "active" part below) from the part of the liver that does not participate in regeneration (the "passive" part). We take into account that the passive part still requires nutrients for maintenance, but do not contribute to biomass production. More and more cells from the passive part get activated until the whole organ contributes to growth. Computationally, this means that at each integration time step we transfer an elementary volume from the passive part to the active part.

We model the reduction of the passive part $V_P$ during growth as:

$$V_P = V_b(1-A) \times \left(\frac{V_J V_b - V_C}{V_J V_b - V_b}\right)^p \quad (10)$$

where $A$ is the fraction of the initial active part, $V_b$ is the initial liver volume; $V_J$ is the relative volume at the joining point, $V_C$ is the total volume of the growing liver, $p$ is a power that allows varying the functional dependence $V_P(V_C)$, choosing different concave and convex shapes. Equation 10 reflects the monotonic increase of the active liver volume. The exponent $p$ accounts for deviations from purely linear increase (when $p=1$). Both parameters $p$ and $A$ are found by fitting to experimental data, including the one before the "joining point". The volumes at the joining points for dog 1 and dog 2, according to our previous computations, are $V_{J1}=1.1576$ and $V_{J2}=1.6918$. Note that equation 10 is constructed in such a way that the passive volume becomes zero when $V_C = V_J V_b$. The active growing part is the complement of the passive part, hence

$$V_A = V_C - V_P \quad (11)$$

The growth equation for a partially growing liver is:

$$pdV(r,d) = K \times (V_A/V_b) \times \left(\frac{R_S}{R_V} - 1\right)dt \quad (12)$$

Equations 10–12 define a complete system of equations required for computing growth curves for partially growing livers. Computed growth curves for the whole growth cycle, including the phase of partial growth, are shown in Fig. 3 for both dogs considered.

The parameters $p$ and $A$ reveal that initially about half of the organ is engaged in proliferation, and that proliferation increases until normal growth is reached. The computed increase of the actively growing part of the liver, defined as $a = (V_b - V_P)/V_b$, is shown in Fig. 3 for both dogs (dashed curves). If the resection is small, and the liver size is close to original, then there is no growth phase in which all hepatocytes are involved in proliferation. In this case, our considerations are valid with regard to the *fraction* of hepatocytes contributing to normal, evolutionarily developed growth as described by the growth equation.

Is the result of identifying $p$ and $A$ from the data unique, or could there be several combinations of $p$ and $A$ that would lead to similar results? We claim that the found values are unique and robust. The reason is that parameter $A$ defines the shape of the *whole* growth curve, while parameter $p$ affects only the shape of the growth curve *before* the joining point. Besides, the shape of the growth curve is very sensitive to the value of $A$, indicating that this parameter has a good identifyability; changing $A$ by few percent increases the difference between the computed growth curve and the experimental data by tens of percent. In fact, $A$ influences the shape and location of the growth curve substantially more than $p$, even during partial growth. We hence first only find the value of $A$ such that we obtain the least diversion of the computed growth curve from the experimental data. Then we identify $p$ to best represent the partial growth phase. This alternating optimization scheme is then iterated until convergence is reached. While the results are very sensitive to the value of $A$, the value of $p$ causes orders of magnitude less change in the diversion from experimental data.

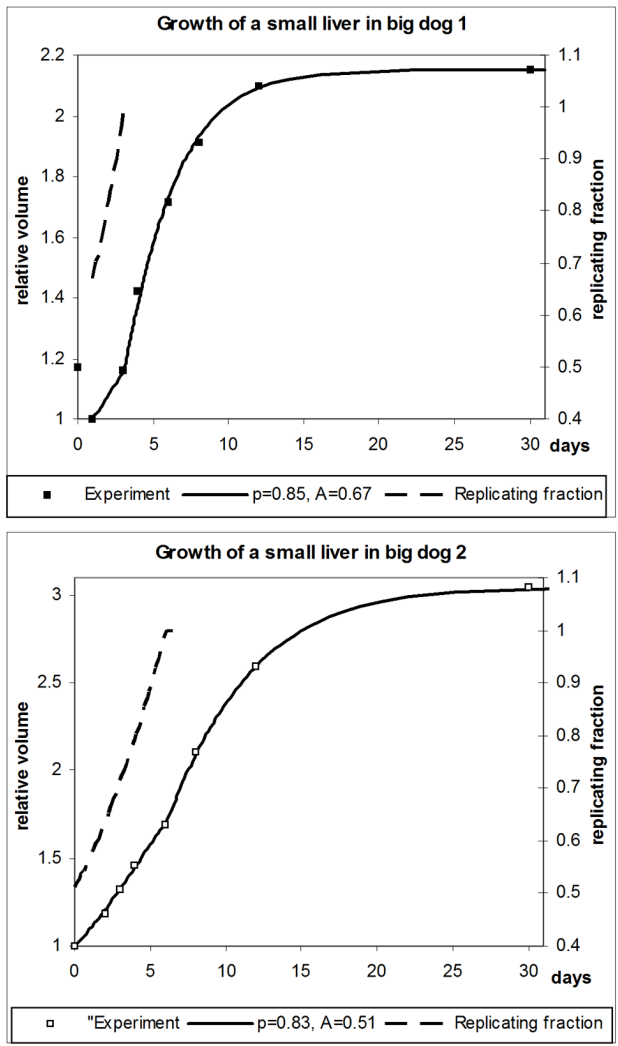

**Figure 3. Partial growth of the liver in two dogs.** Computed growth curves versus experimental data and fractions of replicating hepatocytes relative to normal growth. A - dog 1; B - dog 2.
doi:10.1371/journal.pone.0099275.g003

Overall, the correspondence between the computed growth curves and experimental data can be considered good within the entire growth period.

## Results and discussion for Models of Dog Livers

Based on the computed growth curves, we can make several observations: First, using the proposed method in case of substantial resections, we are able to accurately estimate when partial growth is complete and the entire liver begins to grow (or, in case of small resections, when normal growth begins). This is important information, which previously could not be obtained from observations. In the case of the first dog, whose transplanted liver was larger relative to the final size, this happened after 3 days, while the liver of the second dog began regenerating normally (and in this case apparently entirely because of the small original size) after 6 days.

Although the second dog had a noticeably smaller initial liver size compared to the final size, the fractions of volumes corresponding to the joining points (relative to the final liver size) for both dogs are remarkably close and within overlapping error margins:

$$v_{J1} = V_{J1}/V_{e1} = 1.1576/2.15091 = 0.5382 \pm 0.026$$

$$v_{J2} = V_{J2}/V_{e2} = 1.6918/3.045332 = 0.5555 \pm 0.019$$

Here, we take into account that the error of volume measurement by CT scan is about 5% [16], which translates to errors of $\pm 0.026$ and $\pm 0.019$ for dogs 1 and 2, respectively.

From this, one may conclude that regardless of the initial size of a liver transplant, the liver begins to normally regenerate when its relative size (relative to the final size) is about 54%. Before reaching the joining point, the liver can grow only partially. One possible explanation for this is that a liver, which is smaller than the size corresponding to the joining point, is under stress, and its first priority is to support the functioning of the organism, while less nutritional resources and less hepatocytes can be allocated for growth.

We were also able to quantify another important parameter: the fraction of liver that is involved in regeneration from the very beginning. In the case of the first dog, which had a larger initial liver transplant, this was about $67 \pm 1.8\%$ of the entire liver, while in the smaller liver transplant of the second dog about $51 \pm 1.6\%$ of the initial liver volume contributed to regeneration. (The error was estimated for a 5% change of the average deviation of the computed growth curve from the experimental data). Overall, we see that in both cases a significant portion of the liver is involved in regeneration from the beginning.

The rates of liver growth in both dogs were almost identical. Although we do not have data for smaller fractions for the first dog, because the original size of the liver transplant was larger, we can compare rates of liver growth in the last two days before the joining points. For the first dog the relative increase in volume during these two days was $\Delta v_1 \approx 0.1576$, while for the second dog it was $\Delta v_1 = V_{J2}/V_4 \approx 0.1567$, where $V_4$ is the liver's relative volume on day 4. Given the measurement error of about 2%, this is a remarkable similarity in the rates of liver regeneration. Of course, having results from only two dogs does not allow definitive conclusions.

Lastly, we were able to identify the rate at which the passive part of the liver joins active regeneration. Fig. 3 (dashed lines) shows two convex curves with exponents 0.85 and 0.83 for the first dog and the second dog, respectively. So, the rate at which "passive" liver parts become "active" is similar for both dogs, and it accelerates toward the "joining" point. Both features are physiologically justified, since fundamental mechanisms of liver regeneration should not significantly differ across different specimens, while acceleration of the rate is facilitated by liver growth, which allows devoting progressively more resources to regeneration, while continuing to support the physiological requirements of the organism.

## Modeling Liver Growth in Humans

### 3.1. Geometrical model of a human liver

We further validate our model using experimental data on the growth of transplanted livers in humans from [28,29]. We model human liver geometry as a prism with one edge cut as shown in Fig. 4, based on liver description from [26]. In clinical practice, either the right or left liver lobe is transplanted, leaving the donor with the remaining lobe. In Fig. 4, the right lobe is on the left, and vice versa, since this is how livers are presented in the anatomy literature. In the previous case of dogs, whole livers were transplanted and then grew in size at constant shape. Here, the growing liver changes shape as a single lobe regenerates to a full liver. The geometric form of the liver hence changes during growth. The geometric characteristics of whole livers, as taken from [26,28,29], are given in Table 2.

Using the same consideration as for dog livers, we can also assume that the nutrient influx per unit volume in human livers is constant. This assumption and the parameters from Table 2 constitute a *complete set* of parameters required to compute the growth curve for the model of a human liver. Other parameters, such as graft lengths, can be computed using the formulas below.

For the simulations, we also need the angle α at the prism's base (see Fig. 4), which is defined as:

$$tg\alpha = L/(B_X - B) \tag{13}$$

Let us denote $t = tg\alpha$ for brevity. Then, in the notation of Fig. 4, we can find the surface and volume of the cut prism as:

$$V(B,L,t) = W(BL + L^2/(2t)) \tag{14}$$

$$S(B,L,t) = W(2B + L/t) + WL(1 + \sqrt{1 + 1/t^2}) + L(2B + L/t) \tag{15}$$

The equivalent expressions in terms of the larger base $B_X$ are more convenient for computations of the right lobe:

$$V(B_X,L,t) = W(B_X L - L^2/(2t)) \tag{16}$$

$$S(B_X,L,t) = W(2B_X - L/t) + WL(1 + \sqrt{1 + 1/t^2}) + L(2B_X - L/t) \tag{17}$$

equation 17 is the sum of the areas of all prism faces.

The mass or volume of the graft taken from the donor for transplantation is usually recorded as a fraction of the total size of the donor's liver. We denote this fraction $F$ (fraction). Then, in case of a left-lobe transplant, we find initial length $L_b$ as follows: We rewrite equation 16 as:

Please see Figure 4 in the attachment.

**Figure 4. Geometric model of a human liver.** The boundary plane defines the initial volume of the transplanted lobe. It can be shifted along the direction of arrow *A*.
doi:10.1371/journal.pone.0099275.g004

$$FV(B,L,t) = W(BL + L^2/(2t)) \tag{18}$$

Solving this equation for $L_{bL}$, we find:

$$L_{bL}(B,t,W,F) = -Bt + \sqrt{B^2 t^2 + 2tFV/W} \tag{19}$$

Substituting the volume $V$ from equation 14 into equation 19, we can rewrite it as:

$$L_{bL}(B,t,F) = -Bt + \sqrt{B^2 t^2 + 2FBLt + FL^2}$$

Similarly, we can find the initial length $L_{bR}$ for a transplanted right lobe:

**Table 2.** Geometric parameters of initial liver grafts and whole livers (from [26,28,29]).

| Parameter | Male | Female |
| --- | --- | --- |
| Width of a whole liver (relative units) | 1, 2 | 1, 2 |
| Small base $B$ of a whole liver (in units of width) | 1 | 1 |
| Large base of a whole liver $B_X$ (in units of width) | 3.5 | 3.5 |
| Length of a whole liver (in units of width) | 2.9 | 2.9 |
| Initial volume (percentage of the original donor liver) | 48.5 | 59.6 |
| Final volume (percentage of the original donor liver) | 85.42 | 79.58 |
| Relative final volume (relative to minimum) | 1.7612 | 1.5605 |

doi:10.1371/journal.pone.0099275.t002

$$L_{bR}(B_X,t,F) = B_X t - \sqrt{B_X^2 t^2 - 2FB_X Lt + FL^2} \quad (20)$$

Equations 15–20 uniquely define the shapes of the transplanted and remaining liver grafts.

### 3.2. Growth of the remaining graft in human donors

Reference [28] focused on the growth of the remaining parts of livers in donors, whose safety was a primary goal of that study. Fig. 5A presents results for 27 male donors who had their right lobes removed, so that their left lobes had to regrow to full livers. The data points show average and standard deviation values across all 27 donors. Comparing with female donors (Fig. 5B), the growth phase in male donors lasted longer. There are several plausible explanations for this difference, but in lack of experimental data no sound conclusion can be made. The authors also noted that "Female donors had significantly slower liver regrowth when compared to males at 12 months ($79.8 \pm 9.3\%$ versus $85.6 \pm 8.2\%$)". This result is almost surely due to higher metabolic capacity of female livers required to support pregnancy (this feature has been discovered and discussed in the second article on liver metabolism), so that neither female liver transplants nor liver remnants in females need to grow as big as in males, since their higher metabolic capacity allows supporting metabolic requirements by having smaller size.

Given the inter-patient fluctuations, the computed growth curve from our model corresponds well with the experimental data. Overall, the present model also accurately reproduces the growth dynamics of organs whose geometric shape changes during growth.

### 3.3. Comparison of liver growth rates for the left and right lobes: The effect of geometry

In [29], the authors studied the growth of livers in both donors and recipients for both left-lobe donors and left-lateral-section donors. It was discovered that "Livers of the right lobe donor group regenerated fastest in the donors group…". Our model, which is based on the general growth law, readily explains this effect. Looking at Fig. 4, the *thickest* part of the liver has to regrow during the regeneration of a *left* lobe. When the *right* lobe grows, then the *thinnest* part of the liver (on the right in Fig. 4) has to regrow. Such different growth geometries lead to different values of the growth ratio and consequently to different rates of growth for left and right lobes. Fig. 6 shows the change in value of the growth ratio during growth of right and left lobes when the relative initial volumes are the same (females). We see that the value of the growth ratio for the left lobe is higher than for the right lobe, although all other parameters are the same, except geometry. Since the amount of nutrients available for biomass production is defined by the growth ratio, this means that for the same nutrient supply the amount of biomass produced per unit time is higher in growing left lobes than it is in right lobes, which, indeed, was experimentally observed [29].

The aforementioned difference in growth ratios between right and left lobes creates a difference in the rate of biomass production of several percent. For instance, after 108 days the rates of biomass production between the right and left lobes differ by about 7%.

Note that the discussed difference in values of growth ratios when the relative initial volumes of right and left lobes are the same is not the only factor contributing to different rates of liver growth. Differences in the initial relative volume also influence the value of the growth ratio and consequently the rate of biomass production, as seen for the male cohort. In Fig. 6, the right lobe remnant in males began to grow at a size of about 69% of the final liver volume, while the size of left lobe remnants was about 47% of the final volume. Both remnants grew to about the same final size. As we can see, the value of the growth ratio for the left lobe remnant is *significantly* higher than for the right lobe, which in this case is due to different geometries of right and left lobes, and also due to different initial volumes. (Strictly speaking, different initial volumes also influence geometrical characteristics, but the result of this influence is of lesser value.) The simultaneous action of the two discussed effects explains the observed differences - the faster growth of left lobes in donors that donated right lobes.

## Developing Integral Models

The liver is an important part of any organism. Its working interrelates to other organism systems and organs. On the other hand, it is a separate organ with specialized functions, whose relationship to other organs and systems can be described in terms of inputs and outputs. Indeed, given the multitude of different, often interrelated, factors that affect liver function, its mathematical modeling presents a challenge. In such a situation, it is especially important to provide a robust and adequate modeling structure (including hierarchical relationships) that would incorporate different scales, from molecular mechanisms to the whole organ. In this regard, the proposed model presents a valuable and, in fact, unique development, since the general growth law is applicable at dimensional scales from cellular components to entire organisms, thus providing a universal conceptual approach and the same uniform mathematical apparatus for different scale levels. This way, all meaningful parameters, such as, for instance, nutrient influxes, can be related from the lower scale level to the upper one, up to the integral parameters characterizing the whole organ. Note that exactly the same concept and mathematical framework is entirely applicable to other systems, organs and their subcomponents, which is a consequence of the universal nature of the general growth law [17]. Such a universal approach *significantly* simplifies mathematical modeling of organisms and their constituents.

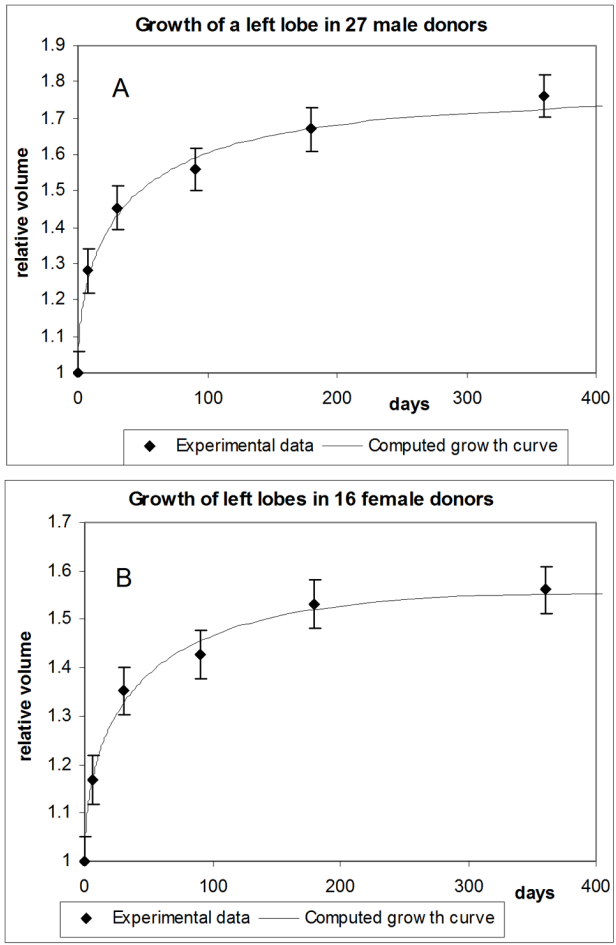

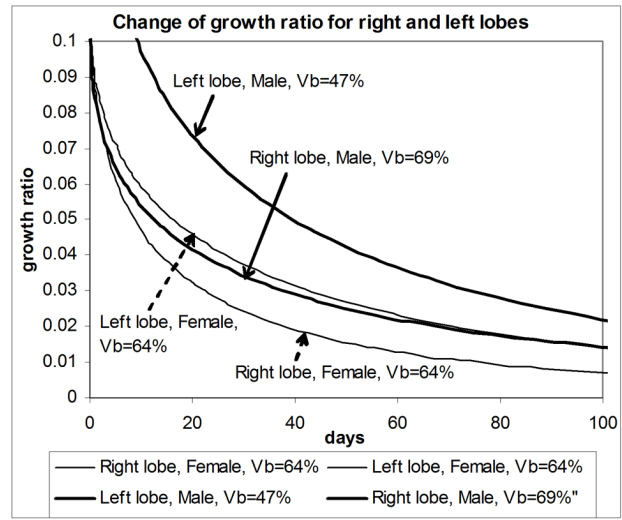

**Figure 6. Change of the growth ratio for growing right and left liver lobes.** The effect is due to changing liver geometry during growth. Scenarios are presented when initial lobe volumes are different and when they are the same.
doi:10.1371/journal.pone.0099275.g006

**Figure 5. Growth of remaining left liver lobes in human donors.** A – male donors; B – female donors.
doi:10.1371/journal.pone.0099275.g005

However, would the above be sufficient to develop an adequate *biophysical* and *biochemical* model of a complex organ? In principle, the answer is yes. The proposed approach resolves several fundamental issues defining success of any modeling, such as model uniqueness, stability, scalability and integrity. In practical applications, much attention is given to biochemical mechanisms because of their importance for medical, pharmaceutical and biotechnological purposes. In this regard, presently the relationship between the integral characteristics, such as nutrient consumption or geometrical form of an organ or cells, and biochemical reactions are very weakly explored, if at all. This is why the "biochemical part of the story" is usually self-enclosed, although it is far from being self-sufficient. In such a situation, the use of the general growth law and models developed on its basis, like the one which we introduced in this work, become of critical importance, since they enable to directly relate integral characteristics, such as nutrient influx and amount of produced biomass, to composition of biochemical reactions and to geometric size and shape. In this arrangement, the amount of produced biomass (which in turn is defined by the growth ratio) is a leading indicator that defines composition of biochemical reactions. This fact is well studied at a cellular level [13,17] with the aid of methods of metabolic flux analysis. However, according to the general growth law, the same is true at the organ level. Thus, we acquire a very important *universal* link between the composition of biochemical reactions, integral nutrient influxes and biomass production at the *organ level*. Of course, further studies are required to realize this potential.

## Model Applications

The introduced model and the obtained results can be applied in different areas of biology and medicine. Real phenomena, by their nature, are multifactorial. One of the advantages of the proposed model is that it provides a general understanding of an organ's growth dynamics in relation to many other factors. In other words, it allows seeing the overall, often dynamically changing, picture. For instance, in liver transplantation the patient's safety and fast recovery are priorities. Although there were successful transplantations when donors were left with only about 30% of their original liver volume [28,29], in other instances, donors with a substantially bigger part of livers died, so it is a combination of different factors that secures positive outcome. In [28], the authors list diverse reasons for rejecting donors, which confirms this fact. So, any additional information can potentially be useful if it is correctly interpreted.

We found that in the case of dogs there is apparently a stable relative size of a regenerating liver, equal to roughly 54% of the grown organ, when the normal growth begins. In case of humans, a similar effect most likely exists, so that finding such a value for people would allow having a reliable *quantitative* parameter related to successful recovery. We were also able to evaluate the percentage of liver mass actively involved in proliferation below this threshold depending on the phase of growth. This is also a valuable parameter which serves as a good indicator of the metabolic stress the liver transplants (or the liver remnants in donors) are subjected to, since at this critical stage of growth the liver has to support both the metabolic needs of an organism and, at the same time, its own growth.

Another discovered result useful for clinical and other applications relates to the close relationship between the size of a growing liver and its biochemical properties. What is even more important, we were able to introduce a quantitative parameter, the growth ratio, which quantifies such a relationship through the amount of produced biomass. In fact, the found relationship unambiguously works in both directions, that is, once we know the current size of a growing liver, we can make predictions about the composition of biochemical reactions. Inversely, once we know certain specific biochemical characteristics, we can evaluate the relative size of a growing liver compared to its final size, which would be a nice noninvasive inexpensive method for controlling the recovery process. Such a possibility is confirmed by observations from [24] with regard to ornithine decarbohylase, whose concentration depends on the phase of growth. Since biochemical reactions do not proceed in isolation, but are tightly interrelated to each other within the same biochemical machinery, this approach looks promising, since knowledge of the content of several substances fundamentally allows restoring the *overall* composition of biochemical reactions.

Close values of biomass increase rates, which we obtained for dogs, present another observation worthy of attention, since if it is valid for people, it allows introducing a quantitative reference value, to which the recovery process may refer to.

Abilities of livers to regenerate depend on their metabolic capacity, which is indirectly evidenced by results obtained in referenced works. We already briefly discussed that the metabolic capacity of female livers is noticeably higher than that of male livers, of which the smaller final size of livers in females [28] is one of the effects. Such a sexual distinction is an important factor to be taken seriously in clinical practice. It means that a female donor can be safely left with a smaller part of liver than a male donor. For male donors, the size of liver remnants is more critical for successful recovery, all other factors being equal.

The mere fact that the liver size and its metabolic capacity interrelate also provides interesting possibilities. Of course, lifestyle influences metabolic requirements, and accordingly affects liver size. However, when all other factors are equal, a smaller liver would mean higher metabolic capacity. So that maybe a small liver is not a so restrictive factor for transplantation purposes, although in the study [28] "inadequate liver volume" contributed to 19.5% of donor rejections.

Of course, the considered examples by no means exhaust possible clinical and other applications of the presented model and obtained results. This is a general model, which is based on a fundamental law of nature, so that it can be used for a very wide range of purposes. In this section, we just scratched the surface discussing examples of possible applications.

## Conclusions

Based on the earlier discovered general growth law, presented in [17], we proposed a macroscopic model for volumetric growth of organs that accounts for quantitative characteristics of growth and for the geometric shape of the organ. We exemplified the use of the resulting model by applying it to modeling growth of transplanted livers and to identifying characteristics of growing livers in dogs and humans. We validated the model by comparison with available experimental data from the literature on growth of liver transplants in dogs and liver grafts and remnants in humans. In the case of dogs, we modeled growth of whole livers, so that we have had a proportional increase of the whole organ, whose shape thus did not change during growth. In the case of humans, we modeled growth of liver grafts obtained from donors as a result of hepatectomy, and liver remnants in donors, so that the liver has been changing its geometric form during growth.

We made the following observations:

1. A dood agreement between experimental data and the theoretically predicted growth curves for growing livers, liver grafts, and liver remnants was discovered.

2. We were able to determine the time point when a liver switches from partial growth to a normal, evolutionarily developed, growth (i.e., the joining point) in dogs. This result can be used for optimizing the size of liver transplants and the fraction of liver left in the donor.

3. The portion of the liver in dogs that participates in regeneration from the very beginning was found.

4. We found the functional dependence of the conversion of "passive" (with regard to growth) liver parts to "active", growing parts in dogs.

5. We discovered apparently stable relationships between the size of a fully grown liver and the time point when the liver switches to normal regeneration (in case of large resections, to a full regeneration).

6. In dogs, the rates of liver growth before the joining point are similar.

7. In humans, the fact that left-lobe liver remnants grow faster than right-lobe remnants is partially due to differences in their geometry. We qualitatively described this effect and found that it may account for about 10% of the difference in growth rates, depending on the initial volume of liver graft relative to the whole liver.

Although we focused on modeling growth of livers, the present method can potentially also be applied to modeling growth of other organs or whole organisms.

Our results show that the growth equation, which is the mathematical representation of the general growth law, is an adequate quantitative and phenomenological tool for many practical applications and theoretical studies. It accurately describes the dynamics of organ growth in quantitative terms, and it allows hypothesizing about the mechanisms underlying many effects observed in experimental studies.

The proposed method can be used for quantitative estimation of the optimal size of liver transplants from the perspective of patient safety and recovery time. The present method also allows optimizing the shape of transplants, and provides quantitative indications for nutrient supply in safe and fast recovery.

A related method, also based on the general growth law, for finding metabolic characteristics of organisms and their constituents (cells, organs, etc.) has been developed and experimentally verified using data on liver and liver transplants. It allows finding rates of nutrient consumption for growth and maintenance, and the total amount of nutrients required for growth. These studies are presented in a second article.

## Acknowledgments

The authors thank A. Y. Shestopaloff for discussions, reviews, and editing efforts, Dr. P. H. Pawlowski (Institute of Biochemistry and Biophysics, Polish Academy of Sciences) for continuous support in the study of the general growth law, and Reviewer for valuable comments.

## Author Contributions

Analyzed the data: YS. Wrote the paper: YS. Editing manuscript: YS IS. Conceiving the idea to study liver growth: YS IS. Designed software used in analysis: YS. Discussions of the results and material: YS IS.

# A Method for Modeling Growth of Organs and Transplants Based on the General Growth Law: Application to the Liver in Dogs and Humans

Yuri K. Shestopaloff[1], Ivo F. Sbalzarini[2]

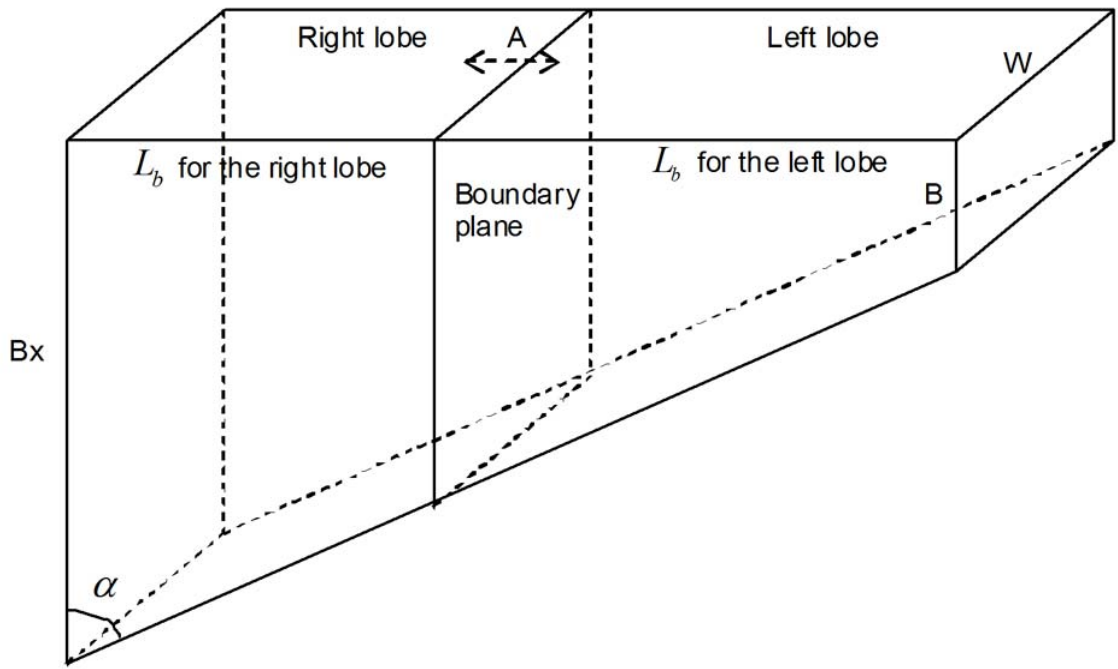

**Figure 4. Geometric model of a human liver.** The boundary plane defines the initial volume of the transplanted lobe. It can be shifted along the direction of arrow *A*.

1 Research and Development Lab, Segmentsoft Inc., Toronto, Ontario, Canada,
2 MOSAIC Group, Center of Systems Biology Dresden (CSBD), Max Planck Institute of Molecular Cell Biology and Genetics, Dresden, Germany